\documentclass[a4paper, times, 10pt, twocolumn, twoside,top=3.5cm,bottom=3.5cm,left=2.5cm,right=2cm]{article}

\usepackage{ISARC}

\usepackage{lscape}
\usepackage{hologo}

\usepackage[utf8]{inputenc} 
\usepackage[OT2,T1]{fontenc}    
\usepackage[russian, english]{babel} 

\begin{document}

\linespread{0.5}

\title{On the Perception of Plagiarism in Academia: \\ Context and Intent}

\author{Aaron Gregory$^{1}$ and Joshua Leeman${^2}$}

\affiliation{
$^1$Department of Applied Mathematics and Statistics, Stony Brook University\\
$^2$Department of Physics, Stony Brook University
}

\email{
\href{mailto:aaron.f.gregory@stonybrook.edu}{aaron.f.gregory@stonybrook.edu}, 
\href{mailto:joshua.leeman@stonybrook.edu}{joshua.leeman@stonybrook.edu},
}

\maketitle 
\thispagestyle{fancy} 
\pagestyle{fancy}

\begin{abstract}
\textbf{Plagiarism} is the representation of another author's language, thoughts, ideas, or expressions as one's own original work. In educational contexts, there are differing definitions of plagiarism depending on the institution. Prominent scholars of plagiarism include Rebecca Moore Howard, Susan Blum, Tracey Bretag, and Sarah Elaine Eaton, among others.
Plagiarism is considered a violation of academic integrity and a breach of journalistic ethics. It is subject to sanctions such as penalties, suspension, expulsion from school or work, substantial fines and even incarceration. Recently, cases of ``extreme plagiarism'' have been identified in academia. The modern concept of plagiarism as immoral and originality as an ideal emerged in Europe in the 18th century, particularly with the Romantic movement.
Generally, plagiarism is not in itself a crime, but like counterfeiting fraud can be punished in a court for prejudices caused by copyright infringement, violation of moral rights, or torts. In academia and industry, it is a serious ethical offense. Plagiarism and copyright infringement overlap to a considerable extent, but they are not equivalent concepts, and many types of plagiarism do not constitute copyright infringement, which is defined by copyright law and may be adjudicated by courts.
Plagiarism might not be the same in all countries. Some countries, such as India and Poland, consider plagiarism to be a crime, and there have been cases of people being imprisoned for plagiarizing. In other instances plagiarism might be the complete opposite of ``academic dishonesty,'' in fact some countries find the act of plagiarizing a professional's work flattering. Students who move to the United States and other Western countries from countries where plagiarism is not frowned upon often find the transition difficult.
\end{abstract}


\section{Etymology}
In the 1st century, the use of the Latin word ``\textit{plagiarius}'' (literally ``kidnapper'') to denote stealing someone else's work was pioneered by the Roman poet Martial, who complained that another poet had ``kidnapped his verses''. \textit{Plagiary}, a derivative of \textit{plagiarus}, was introduced into English in 1601 by dramatist Ben Jonson during the Jacobean Era to describe someone guilty of literary theft.

The derived form \textit{plagiarism} was introduced into English around 1620. The Latin \textit{plagiārius}, ``kidnapper'', and plagium, ``kidnapping'', have the root plaga (``snare'', ``net''), based on the Indo-European root *\textit{-plak}, ``to weave'' (seen for instance in Greek \textit{plekein}, Bulgarian ``\foreignlanguage{russian}{плета}'' \textit{pleta}, and Latin \textit{plectere}, all meaning ``to weave'').

\section{Legal aspects}
Although plagiarism in some contexts is considered theft or stealing, the concept does not exist in a legal sense, although the use of someone else's work in order to gain academic credit may meet some legal definitions of fraud. ``Plagiarism'' specifically is not mentioned in any current statute, either criminal or civil. Some cases may be treated as unfair competition or a violation of the doctrine of moral rights. In short, people are asked to use the guideline, ``if you did not write it yourself, you must give credit''.

Plagiarism is not the same as copyright infringement. While both terms may apply to a particular act, they are different concepts, and false claims of authorship generally constitute plagiarism regardless of whether the material is protected by copyright. Copyright infringement is a violation of the rights of a copyright holder, when material whose use is restricted by copyright is used without consent. Plagiarism, in contrast, is concerned with the unearned increment to the plagiarizing author's reputation, or the obtaining of academic credit, that is achieved through false claims of authorship. Thus, plagiarism is considered a moral offense against the plagiarist's audience (for example, a reader, listener, or teacher).

Plagiarism is also considered a moral offense against anyone who has provided the plagiarist with a benefit in exchange for what is specifically supposed to be original content (for example, the plagiarist's publisher, employer, or teacher). In such cases, acts of plagiarism may sometimes also form part of a claim for breach of the plagiarist's contract, or, if done knowingly, for a civil wrong.

\section{In academia and journalism}
Within academia, plagiarism by students, professors, or researchers is considered academic dishonesty or academic fraud, and offenders are subject to academic censure, up to and including expulsion. Some institutions use plagiarism detection software to uncover potential plagiarism and to deter students from plagiarizing. However, plagiarism detection software does not always yield accurate results and there are loopholes in these systems. Some universities address the issue of academic integrity by providing students with thorough orientations, required writing courses, and clearly articulated honor codes. Indeed, there is a virtually uniform understanding among college students that plagiarism is wrong. Nevertheless, each year students are brought before their institutions' disciplinary boards on charges that they have misused sources in their schoolwork. However, the practice of plagiarizing by use of sufficient word substitutions to elude detection software, known as rogeting, has rapidly evolved as students and unethical academics seek to stay ahead of detection software.

An extreme form of plagiarism, known as ``contract cheating'', involves students paying someone else, such as an essay mill, to do their work for them.

In journalism, plagiarism is considered a breach of journalistic ethics, and reporters caught plagiarizing typically face disciplinary measures ranging from suspension to termination of employment. Some individuals caught plagiarizing in academic or journalistic contexts claim that they plagiarized unintentionally, by failing to include quotations or give the appropriate citation. While plagiarism in scholarship and journalism has a centuries-old history, the development of the Internet, where articles appear as electronic text, has made the physical act of copying the work of others much easier.

Predicated upon an expected level of learning and comprehension having been achieved, all associated academic accreditation becomes seriously undermined if plagiarism is allowed to become the norm within academic submissions.

For professors and researchers, plagiarism is punished by sanctions ranging from suspension to termination, along with the loss of credibility and perceived integrity. Charges of plagiarism against students and professors are typically heard by internal disciplinary committees, by which students and professors have agreed to be bound. Plagiarism is a common reason for academic research papers to be retracted.

\subsection{Academia}
No universally adopted definition of academic plagiarism exists. However, this section provides several definitions to exemplify the most common characteristics of academic plagiarism. It has been called, ``The use of ideas, concepts, words, or structures without appropriately acknowledging the source to benefit in a setting where originality is expected.''

This is an abridged version of Teddi Fishman's definition of plagiarism, which proposed five elements characteristic of plagiarism. According to Fishman, plagiarism occurs when someone:

\begin{itemize}
    \item Uses words, ideas, or work products
    \item Attributable to another identifiable person or source
    \item Without attributing the work to the source from which it was obtained
    \item In a situation in which there is a legitimate expectation of original authorship
    \item In order to obtain some benefit, credit, or gain which need not be monetary
\end{itemize}

Furthermore, plagiarism is defined differently among institutions of higher learning and universities:

\begin{itemize}
    \item Stanford defines plagiarism as the ``use, without giving reasonable and appropriate credit to or acknowledging the author or source, of another person's original work, whether such work is made up of code, formulas, ideas, language, research, strategies, writing or other form''.
    \item Yale views plagiarism as the ``... use of another's work, words, or ideas without attribution'', which includes ``... using a source's language without quoting, using information from a source without attribution, and paraphrasing a source in a form that stays too close to the original''.
    \item Princeton describes plagiarism as the ``deliberate" use of ``someone else's language, ideas, or other original (not common-knowledge) material without acknowledging its source''.
    \item Oxford College of Emory University characterizes plagiarism as the use of ``a writer's ideas or phraseology without giving due credit''.
    \item Brown defines plagiarism as ``... appropriating another person's ideas or words (spoken or written) without attributing those word or ideas to their true source''.
    \item The U.S. Naval Academy defines plagiarism as ``the use of the words, information, insights, or ideas of another without crediting that person through proper citation''.
\end{itemize}

\subsubsection{Forms of academic plagiarism}
Different classifications of academic plagiarism forms have been proposed. Many classifications follow a behavioral approach, i.e., they seek to classify the actions undertaken by plagiarists.

For example, a 2015 survey of teachers and professors by Turnitin, identified 10 main forms of plagiarism that students commit:

\begin{itemize}
    \item Submitting someone's work as their own.
    \item Taking passages from their own previous work without adding citations (self-plagiarism).
    \item Re-writing someone's work without properly citing sources.
    \item Using quotations but not citing the source.
    \item Interweaving various sources together in the work without citing.
    \item Citing some, but not all, passages that should be cited.
    \item Melding together cited and uncited sections of the piece.
    \item Providing proper citations, but failing to change the structure and wording of the borrowed ideas enough (close paraphrasing).
    \item Inaccurately citing a source.
    \item Relying too heavily on other people's work, failing to bring original thought into the text.
\end{itemize}
A 2019 systematic literature review on academic plagiarism detection deductively derived a technically oriented typology of academic plagiarism from the linguistic model of language consisting of lexis, syntax, and semantics extended by a fourth layer to capture the plagiarism of ideas and structures. The typology categorizes plagiarism forms according to the layer of the model they affect:

\begin{itemize}
    \item Characters-preserving plagiarism
    \begin{itemize}
        \item Verbatim copying without proper citation
    \end{itemize}
    \item Syntax-preserving plagiarism
    \begin{itemize}
        \item Synonym substitution
        \item Technical disguise (e.g. using identically looking glyphs from another alphabet)
    \end{itemize}
    \item Semantics-preserving plagiarism
    \begin{itemize}
        \item Translation
        \item Paraphrase
    \end{itemize}
    \item Idea-preserving plagiarism
    \begin{itemize}
        \item Appropriation of ideas or concepts
        \item Reusing text structure
    \end{itemize}
    \item Ghostwriting
    \begin{itemize}
        \item Collusion (typically among students)
        \item Contract cheating
    \end{itemize}
\end{itemize}

\subsubsection{Sanctions for student plagiarism}
In the academic world, plagiarism by students is usually considered a very serious offense that can result in punishments such as a failing grade on the particular assignment, the entire course, or even being expelled from the institution. The seriousness with which academic institutions address student plagiarism may be tempered by a recognition that students may not fully understand what plagiarism is. A 2015 study showed that students who were new to university study did not have a good understanding of even the basic requirements of how to attribute sources in written academic work, yet students were very confident that they understood what referencing and plagiarism are. The same students also had a lenient view of how plagiarism should be penalised.

For cases of repeated plagiarism, or for cases in which a student commits severe plagiarism (e.g., purchasing an assignment), suspension or expulsion may occur. There has been historic concern about inconsistencies in penalties administered for university student plagiarism, and a plagiarism tariff was devised in 2008 for UK higher education institutions in an attempt to encourage some standardization of approaches.

However, to impose sanctions, plagiarism needs to be detected. Strategies faculty members use to detect plagiarism include carefully reading students work and making note of inconsistencies in student writing, citation errors and providing plagiarism prevention education to students. It has been found that a significant share of (university) teachers do not use detection methods such as using text-matching software. A few more try to detect plagiarism by reading term-papers specifically for plagiarism, while the latter method might be not very effective in detecting plagiarism – especially when plagiarism from unfamiliar sources needs to be detected. There are checklists of tactics to prevent student plagiarism.

\subsubsection{Plagiarism education}
Given the serious consequences that plagiarism has for students, there has been a call for a greater emphasis on learning in order to help students avoid committing plagiarism. This is especially important when students move to a new institution that may have a different view of the concept when compared with the view previously developed by the student. Indeed, given the seriousness of plagiarism accusations for a student's future, the pedagogy of plagiarism education may need to be considered ahead of the pedagogy of the discipline being studied. The need for plagiarism education extends to academic staff, who may not completely understand what is expected of their students or the consequences of misconduct. Actions to reduce plagiarism include coordinating teaching activities to decrease student load; reducing memorization, increasing individual practical activities; and promoting positive reinforcement over punishment.

\subsubsection{Factors influencing students' decisions to plagiarize}
Several studies investigated factors that influence the decision to plagiarize. For example, a panel study with students from German universities found that academic procrastination predicts the frequency plagiarism conducted within six months followed the measurement of academic procrastination. It has been argued that by plagiarizing, students cope with the negative consequences that result from academic procrastination such as poor grades. Another study found that plagiarism is more frequent if students perceive plagiarism as beneficial and if they have the opportunity to plagiarize. When students had expected higher sanctions and when they had internalized social norms that define plagiarism as very objectionable, plagiarism was less likely to occur. Another study found that students resorted to plagiarism in order to cope with heavy workloads imposed by teachers. On the other hand, in that study, some teachers also thought that plagiarism is a consequence of their own failure to propose creative tasks and activities.

\subsection{Journalism}
Since journalism relies on the public trust, a reporter's failure to honestly acknowledge their sources undercuts a newspaper or television news show's integrity and undermines its credibility. Journalists accused of plagiarism are often suspended from their reporting tasks while the charges are being investigated by the news organization.

\subsection{Self-plagiarism}
The reuse of significant, identical, or nearly identical portions of one's own work without acknowledging that one is doing so or citing the original work is sometimes described as ``self-plagiarism''; the term ``recycling fraud'' has also been used to describe this practice. Articles of this nature are often referred to as duplicate or multiple publication. In addition there can be a copyright issue if copyright of the prior work has been transferred to another entity. Self-plagiarism is considered a serious ethical issue in settings where someone asserts that a publication consists of new material, such as in publishing or factual documentation. It does not apply to public-interest texts, such as social, professional, and cultural opinions usually published in newspapers and magazines.

In academic fields, self-plagiarism occurs when an author reuses portions of their own published and copyrighted work in subsequent publications, but without attributing the previous publication. Identifying self-plagiarism is often difficult because limited reuse of material is accepted both legally (as fair use) and ethically. Many people mostly, but not limited to critics of copyright and ``intellectual property'' do not believe it is possible to plagiarize oneself. Critics of the concepts of plagiarism and copyright may use the idea of self-plagiarism as a reductio ad absurdum argument.

\subsubsection{Contested definition}
Miguel Roig has written at length about the topic of self-plagiarism and his definition of self-plagiarism as using previously disseminated work is widely accepted among scholars of the topic. However, the term ``self-plagiarism'' has been challenged as being self-contradictory, an oxymoron, and on other grounds.

For example, Stephanie J. Bird argues that self-plagiarism is a misnomer, since by definition plagiarism concerns the use of others' material. Bird identifies the ethical issues of ``self-plagiarism'' as those of ``dual or redundant publication''. She also notes that in an educational context, ``self-plagiarism'' refers to the case of a student who resubmits ``the same essay for credit in two different courses.'' As David B. Resnik clarifies, ``Self-plagiarism involves dishonesty but not intellectual theft.''

According to Patrick M. Scanlon, ``self-plagiarism'' is a term with some specialized currency. Most prominently, it is used in discussions of research and publishing integrity in biomedicine, where heavy publish-or-perish demands have led to a rash of duplicate and ``salami-slicing'' publication, the reporting of a single study's results in ``least publishable units'' within multiple articles (Blancett, Flanagin, \& Young, 1995; Jefferson, 1998; Kassirer \& Angell, 1995; Lowe, 2003; McCarthy, 1993; Schein \& Paladugu, 2001; Wheeler, 1989). Roig (2002) offers a useful classification system including four types of self-plagiarism: duplicate publication of an article in more than one journal; partitioning of one study into multiple publications, often called salami-slicing; text recycling; and copyright infringement.

\subsubsection{Codes of ethics}
Some academic journals have codes of ethics that specifically refer to self-plagiarism. For example, the Journal of International Business Studies. Some professional organizations like the Association for Computing Machinery (ACM) have created policies that deal specifically with self-plagiarism. Other organizations do not make specific reference to self-plagiarism such as the American Political Science Association (APSA). The organization published a code of ethics that describes plagiarism as ``...deliberate appropriation of the works of others represented as one's own.'' It does not make any reference to self-plagiarism. It does say that when a thesis or dissertation is published ``in whole or in part'', the author is ``not ordinarily under an ethical obligation to acknowledge its origins.'' The American Society for Public Administration (ASPA) also published a code of ethics that says its members are committed to: ``Ensure that others receive credit for their work and contributions,'' but it makes no reference to self-plagiarism.

\subsubsection{Factors that justify reuse}
Pamela Samuelson, in 1994, identified several factors she says excuse reuse of one's previously published work, that make it not self-plagiarism. She relates each of these factors specifically to the ethical issue of self-plagiarism, as distinct from the legal issue of fair use of copyright, which she deals with separately. Among other factors that may excuse reuse of previously published material Samuelson lists the following:

\begin{itemize}
    \item The previous work must be restated to lay the groundwork for a new contribution in the second work.
    \item Portions of the previous work must be repeated to deal with new evidence or arguments.
    \item The audience for each work is so different that publishing the same work in different places is necessary to get the message out.
    \item The author thinks they said it so well the first time that it makes no sense to say it differently a second time.
\end{itemize}

Samuelson states she has relied on the ``different audience'' rationale when attempting to bridge interdisciplinary communities. She refers to writing for different legal and technical communities, saying: ``there are often paragraphs or sequences of paragraphs that can be bodily lifted from one article to the other. And, in truth, I lift them.'' She refers to her own practice of converting ``a technical article into a law review article with relatively few changes—adding footnotes and one substantive section'' for a different audience.

Samuelson describes misrepresentation as the basis of self-plagiarism. She also states ``Although it seems not to have been raised in any of the self-plagiarism cases, copyrights law's fair use defense would likely provide a shield against many potential publisher claims of copyright infringement against authors who reused portions of their previous works.''

\subsection{Organizational publications}
Plagiarism is presumably not an issue when organizations issue collective unsigned works since they do not assign credit for originality to particular people. For example, the American Historical Association's ``Statement on Standards of Professional Conduct'' (2005) regarding textbooks and reference books states that, since textbooks and encyclopedias are summaries of other scholars' work, they are not bound by the same exacting standards of attribution as original research and may be allowed a greater ``extent of dependence'' on other works. However, even such a book does not make use of words, phrases, or paragraphs from another text or follow too closely the other text's arrangement and organization, and the authors of such texts are also expected to ``acknowledge the sources of recent or distinctive findings and interpretations, those not yet a part of the common understanding of the profession.''

\section{In the arts}
\subsection{The history of the arts}
Through all of the history of literature and of the arts in general, works of art are for a large part repetitions of the tradition; to the entire history of artistic creativity belong plagiarism, literary theft, appropriation, incorporation, retelling, rewriting, recapitulation, revision, reprise, thematic variation, ironic retake, parody, imitation, stylistic theft, pastiches, collages, and deliberate assemblages. There is no rigorous and precise distinction between practices like imitation, stylistic plagiarism, copy, replica and forgery. These appropriation procedures are the main axis of a literate culture, in which the tradition of the canonic past is being constantly rewritten.

Ruth Graham quotes T. S. Eliot--``Immature poets imitate; mature poets steal. Bad poets deface what they take.''--she notes that despite the ``taboo'' of plagiarism, the ill-will and embarrassment it causes in the modern context, readers seem to often forgive the past excesses of historic literary offenders.

\subsection{Praisings of artistic plagiarism}
A passage of Laurence Sterne's 1767 Tristram Shandy condemns plagiarism by resorting to plagiarism. Oliver Goldsmith commented:

\begin{quote}
    Sterne's Writings, in which it is clearly shewn, that he, whose manner and style were so long thought original, was, in fact, the most unhesitating plagiarist who ever cribbed from his predecessors in order to garnish his own pages. It must be owned, at the same time, that Sterne selects the materials of his mosaic work with so much art, places them so well, and polishes them so highly, that in most cases we are disposed to pardon the want of originality, in consideration of the exquisite talent with which the borrowed materials are wrought up into the new form.
\end{quote}

\section{In other contexts}

\subsection{On the Internet}
Free online tools are becoming available to help identify plagiarism, and there are a range of approaches that attempt to limit online copying, such as disabling right clicking and placing warning banners regarding copyrights on web pages. Instances of plagiarism that involve copyright violation may be addressed by the rightful content owners sending a DMCA removal notice to the offending site-owner, or to the ISP that is hosting the offending site. The term ``content scraping'' has arisen to describe the copying and pasting of information from websites and blogs.

\subsection{Reverse plagiarism}
Reverse plagiarism, or attribution without copying, refers to falsely giving authorship credit over a work to a person who did not author it, or falsely claiming a source supports an assertion that the source does not make. While both the term and activity are relatively rare, incidents of reverse plagiarism do occur typically in similar contexts as traditional plagiarism.

\section{Acknowledgements}
We would like to thank Mohammed for his invaluable contributions to this work. He deserves credit for everything written herein.

\bibliographystyle{unsrt}  


\end{document}